\documentclass[12pt]{reportj}
\usepackage{deluxetablej}
\usepackage{hyperref} 
\makeatletter
\def\@to{to}
\makeatother
\usepackage{times}
\usepackage{comment}

\usepackage{graphicx}
\usepackage{xcolor}
\usepackage{tocloft}
\usepackage{threeparttable}
\usepackage{fancyheadings}
\emergencystretch=\maxdimen
\usepackage{datetime}
\newdateformat{ddmonthyyyy}{\THEDAY\ \monthname[\THEMONTH]\ \THEYEAR}

\renewcommand\thesection{\arabic{section}}
\renewcommand\thesubsection{\thesection.\arabic{subsection}}

\def\ssection#1{\setcounter{subsection}{0} \refstepcounter{section} \section*{\hbox to \hsize{\large\bf \arabic{section}. #1\hfill }}\label{sec} \addcontentsline{toc}{section}{\arabic{section}. #1}}
\def\ssubsection#1{\setcounter{subsubsection}{0} \refstepcounter{subsection}\subsection*{\hbox to \hsize{\normalsize\bfseries\itshape \arabic{section}.\arabic{subsection} #1\hfill}}\label{subsec} \addcontentsline{toc}{subsection}{\arabic{section}.\arabic{subsection} #1}}
\def\ssubsubsection#1{\refstepcounter{subsubsection}\subsection*{\hbox to \hsize{\normalsize\it \arabic{section}.\arabic{subsection}.\arabic{subsubsection} #1\hfill}}\label{subsubsec} \addcontentsline{toc}{subsubsection}{\arabic{section}.\arabic{subsection}.\arabic{subsubsection} #1}}

\def\ssectionstar#1{\section*{\hbox to \hsize{\large\bf #1\hfill}} \addcontentsline{toc}{section}{#1}}
\def\ssubsectionstar#1{\subsection*{\hbox to \hsize{\normalsize\bfseries\itshape #1\hfill}} \addcontentsline{toc}{subsection}{#1}}
\def\ssubsubsectionstar#1{\subsection*{\hbox to \hsize{\normalsize\it  #1\hfill}} \addcontentsline{toc}{subsection}{#1}}

\renewcommand{\cftaftertoctitle}{%
\mbox{}\hfill{\normalfont Page}}
\begin{document}

~\\

\vspace{-2.4cm}
\noindent\includegraphics*[width=0.295\linewidth]{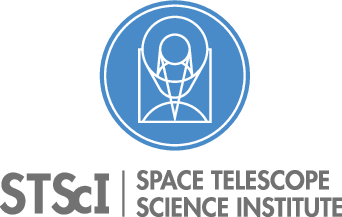}

\vspace{-0.4cm}

\begin{flushright}
    {\bf Instrument Science Report STIS 2026-02}
    
    \vspace{1.1cm}
    
    {\bf\Huge Barycentric Corrections for HST/STIS Data}
    
    \rule{0.25\linewidth}{0.5pt}
    
    \vspace{0.5cm}
    
    Joshua D. Lothringer$^1$, Joleen K. Carlberg$^1$, and Sean Lockwood$^1$
    \linebreak
    \newline
    \footnotesize{$^1$ Space Telescope Science Institute, Baltimore, MD\\}
    
    \vspace{0.5cm}
    
     \ddmonthyyyy{28 May 2026}
\end{flushright}

\vspace{0.1cm}

\noindent\rule{\linewidth}{1.0pt}
\noindent{\bf A{\footnotesize BSTRACT}}

{\it \noindent We describe \texttt{stistools.barycentric\_correction}, a new \texttt{Python} utility for calculating barycentric timing corrections for HST/STIS observations. This tool replaces the deprecated \texttt{stsdas.hst\_calib.stis.odelaytime} IRAF function that was previously used for HST barycentric corrections. Our new utility uses \texttt{astropy} for conversion between time formats and standards and introduces a new way to calculate HST's position through JPL Horizons, replacing the need to download separate HST orbital ephemeris files. Here, we describe the methods used in the new utility, the tests that were carried out to verify its accuracy, and explain some of the complexities involved in determining light travel times to accuracies down to a millisecond for HST. We also summarize the current understanding of the absolute accuracy of STIS time stamps.}

\vspace{-0.1cm}
\noindent\rule{\linewidth}{1.0pt}

\renewcommand{\cftaftertoctitle}{\thispagestyle{fancy}}
\tableofcontents



\vspace{-0.3cm}
\ssection{Introduction}\label{sec:Introduction} 

Because of the finite velocity of light, the time at which events are recorded by telescopes here on Earth or elsewhere need to be translated to a common reference. This common reference point is often taken to be the Solar System's barycenter (SSB), i.e., the center of mass of the Solar System, though occasionally a heliocentric reference frame is also used. In other words, the timing of events like an exoplanet transit or a pulsar's pulse time are converted to the time at which they would be observed at the SSB. This is effectively done by calculating the difference between light's travel time to the observer versus the SSB, called R{\o}mer delay. If the observer is in a gravitational well, relativistic corrections like Einstein delay and Shapiro delay can also be relevant on millisecond timescales. 

This barycentric correction depends on both the observer's location relative to the Sun as well as the observed object's position on the sky. For example, an object observed at opposition to the Sun will have a barycentric correction of about +8.3 minutes because it would take that light an additional 8.3 minutes to travel between the Earth and the SSB. On the other hand, if an object was in conjunction with the Sun, then the barycentric correction would be $-$8.3 minutes since the light would have reached the SSB about 8.3 minutes before reaching Earth.

For some astronomical events and science cases with the \textit{Hubble Space Telescope} (HST), the time at which an event occurs can be known to an accuracy of several seconds without affecting the scientific interpretation. However, some applications like pulsar timing requires timing precision on the order of milliseconds. On such small timescales, positions and velocities of objects generally need to be known at very high accuracy (in astronomical terms): light travels about 300 km in just 1 ms, so the position of an observer must be known to accuracy of better than a few hundred km. Earth's orbital velocity around the Sun is about 30 km/s and the orbital velocity of a satellite around Earth at HST's altitude of about 475~km is about 7.6 km/s. In other words, the position of both Earth and HST need to be calculated accurately every few seconds to enable 1 ms barycentric corrections. Earth's ephemeris can be calculated trivially from built-in \texttt{astropy} functions, but HST's location requires up-to-date information due to its changing orbit. Additionally, the target coordinates must also be accurate: an error of just 1 arcminute in position on the sky results in a timing error of up to 0.28 s in the barycentric correction (Eastman et al. 2010). 

In addition to accurate position and velocity, care must also be taken to ensure sufficient precision in time stamps and consistent time reference frames. A time format describes how a given time is represented. For example, the launch time of the \textit{Hubble Space Telescope} aboard Space Shuttle \textit{Discovery} (April 24, 1990, 12:33:51 UTC) can be expressed as a Julian Date (2448006.02350694), Modified Julian Date (48005.5235069444), ISO 8601 date (1990-04-24 12:33:51), Unix (640960373.816), and GPS (324995579.816). Formats can generally be converted with ease. Sometimes formats can be defined in specific reference frames. For example BJD and HJD are the Julian date at the Barycenter and Heliocenter, respectively.

One also needs to define the time standard for a given timestamp. Time standards (or sometimes time scales, as in \texttt{astropy.time}) are different ways for defining the rate at which time passes. Common time standards include International Atomic Time (TAI, defined in 1967 based on the hyperfine transitions in the ground state of the $^{133}$Cs atom), Terrestrial Time (TT, TAI+32.184 seconds), and Coordinated Universal Time (UTC, TAI+37 leap seconds). Without quoting the time standard used for a given time, a quoted time could be biased by over a minute. See Eastman et al.~(2010) for a comprehensive explanation of various time standards and their origins.

IAU Resolution A4 (1991) formally advocated for the use of Barycentric Julian Date in the Barycentric Dynamical Time standard (BJD$_{\rm TDB}$) for astronomical time-keeping, as explained in Eastman et al.~(2010). BJD$_{\rm TDB}$ is the Julian date at the SSB, corrected for R{\o}mer, Shapiro, and Einstein delay (see Equation~\ref{eq:tdb}). Measuring relative to the barycenter is preferable to measuring relative to the Sun because of the Sun's acceleration from the gravity of the planets, primarily Jupiter and Saturn. With HST often quoted in UTC, additional clock corrections are necessary to convert to the TDB system.

The IRAF task \texttt{odelaytime} was the original method used to calculate barycentric corrections for HST/STIS data and apply them to tag, raw, flt, and x1d files. With IRAF's deprecation, the HST community has been without a bespoke barycentric correction tool for years. Here, we introduce a new utility, \texttt{stistools.barycentric\_correction} to replace the lost functionality of \texttt{odelaytime} with more modern methods and more accurate orbital ephemerides. In what follows, we describe the methods used in \texttt{stistools.barycentric\_correction}, as well as some tests to show its accuracy.

\ssection{Calculating Positions}

\ssubsection{Reference Frames}

\begin{figure}[h]
    \centering
    \includegraphics[width=1\linewidth]{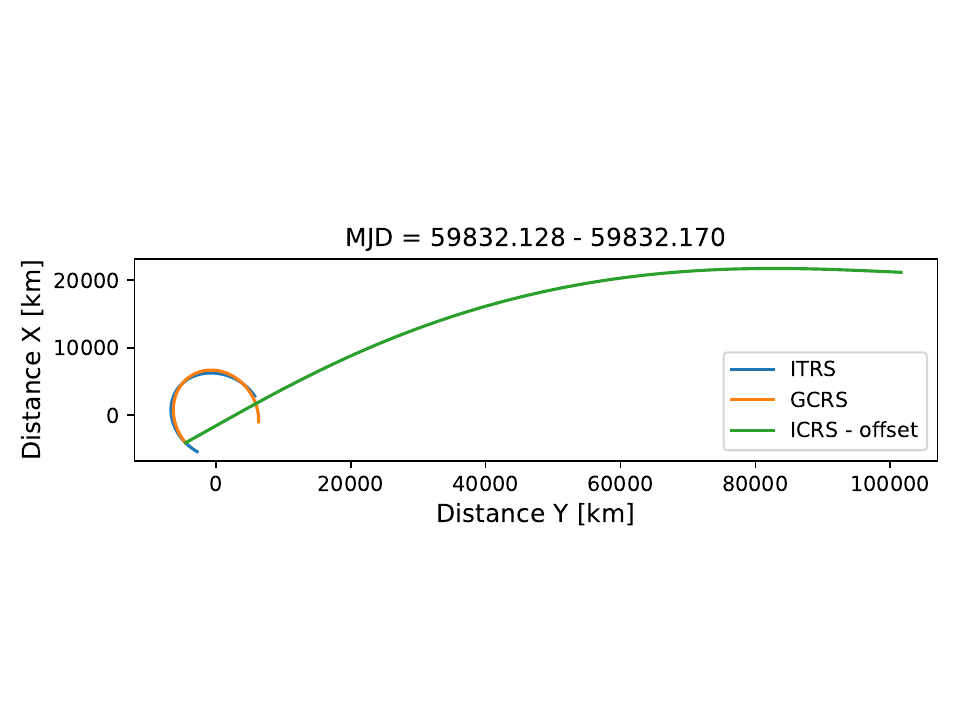}
    \vspace{-80pt}
    \caption{The position of HST over the course of 1 hour on MJD = 59832 in three different reference systems. For visual comparison, an offset equal to the starting position of Earth is subtracted from the ICRS HST position. Additionally, the X and Y axes are swapped for aesthetic reasons.}
    \label{fig:coords}
\end{figure}

Computing time delays to high accuracy and precision also necessitates knowing the relative position of objects (here the SSB, Earth, and HST) to high accuracy and precision. For millisecond timing precision, these positions need to be known to within about a hundred kilometers, though we can often do much better than that. We must therefore pay close attention to the reference frame in which a given position is defined. A reference \textit{frame} refers to a realization (``a catalogue of the adopted coordinates of a set of reference objects") within a reference \textit{system} of procedures, models, and constants (Wilkins 1990). Different methods used to query the position of the Earth and/or HST utilize three different reference systems described below. Figure~\ref{fig:coords} shows the position of HST over the course of an hour in the three different systems.

\textbf{ICRS (International Celestial Reference System):} This system is the standard reference system for astronomy as defined by the IAU, defined using the barycenter of the Solar System as the origin with axes consistent with J2000 declination and right ascension fixed relative to distant quasars.\footnote{\url{https://aa.usno.navy.mil/faq/ICRS_doc}} The latest realization of this system, ICRF3, was adopted in 2019 (Charlot et al. 2020).

\textbf{GCRS (Geocentric Celestial Reference System):} This system is similar to ICRS, but it is defined with the origin at Earth's geocenter.\footnote{\url{https://web.archive.org/web/20101215225721/https://www.iau.org/static/resolutions/IAU2000_French.pdf\#page=5}} Importantly, it does not co-rotate with the Earth and is therefore convenient for defining the position of satellites.

\textbf{ITRS (International Terrestrial Reference System):} A standard Earth-centered, Earth-fixed (i.e., geostationary) reference system that co-rotates with Earth such that latitude and longitude can be meaningfully specified. The system is geocentric (i.e., centered at Earth's center of mass), but some tools like \texttt{barycorrpy} use a similar frame, but defined as geodedic, i.e., defined at a reference ellipse representing Earth's surface, rather than geocentric.

\lhead{}
\rhead{}
\cfoot{\rm {\hspace{-1.9cm} Instrument Science Report STIS 2026-02(v1) Page \thepage}}

\vspace{-0.3cm}

\ssubsection{Calculating Earth's Position}\label{sec:Earthpos}

A first step in calculating the Barycentric correction is to calculate Earth's position. In the IRAF task \texttt{odelaytime}, this was tabulated in a file called \texttt{de200.fits}. ``DE200" refers to a JPL Planetary and Lunar ephemeris used to calculate the position of Earth created in September 1981.\footnote{\url{https://ssd.jpl.nasa.gov/planets/eph_export.html}} With no update since then, \texttt{odelaytime}'s position of Earth is now well out-of-date. The file contained the instantaneous position and velocity of Earth \textit{every two days} from December 14, 1949 00:00:00 to January 02, 2050 00:00:00. No time standard is given for these times, resulting in an uncharacterized error up to a minute. Additionally, with an unknown creation date, anywhere from 0 to 18 leap seconds would need to be applied to the times to update to the present definition of UTC. Lastly, with time sampling every other day, interpolation errors are likely to become important when Earth's orbital position needs to be known accurately on second timescales for millisecond barycentric correction precision. 

\begin{figure}[h]
    \centering
    \includegraphics[width=1\linewidth]{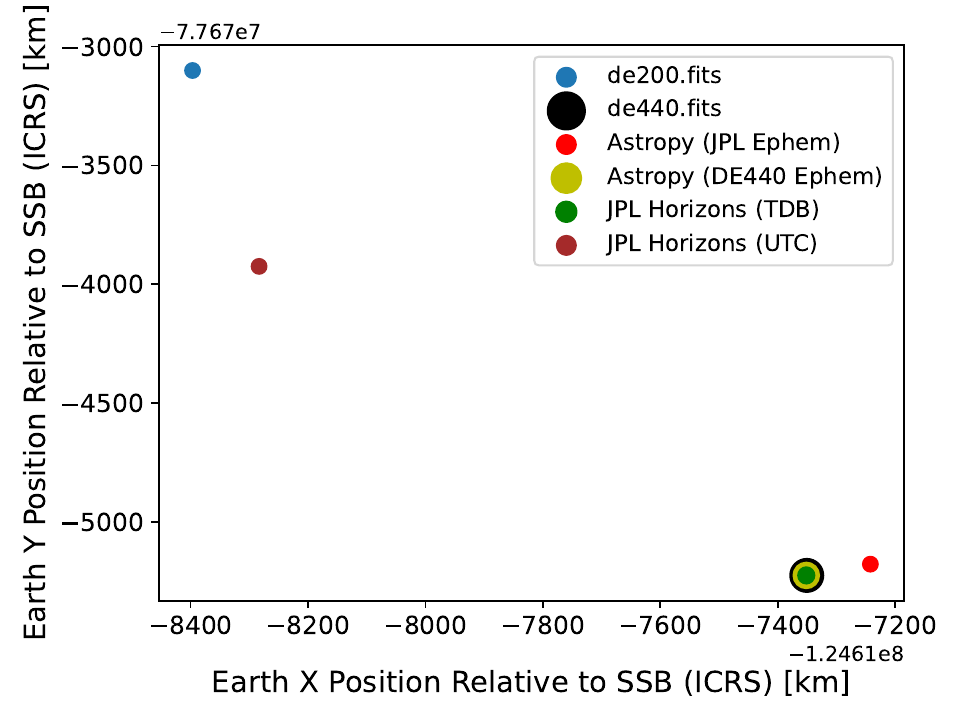}
    \caption{The position of Earth relative to the SSB at the time of HST's launch (April 24, 1990, 12:33:51 UTC) in Cartesian coordinates with different methods. The two fits files refer to the Earth epehemeris file used by our undelivered Python port of \texttt{odelaytime}. Astropy refers to using \texttt{astropy.coordinates.get\_body\_barycentric} with different ephemerides. JPL Horizons uses \texttt{astroquery.jplhorizons} queried at the same MJD but with two different time systems (TDB and UTC) to show their resulting offset. All queries use TDB except the JPL Horizons UTC example.}
    \label{fig:earth}
\end{figure}

Figure~\ref{fig:earth} shows the predicted the position of Earth's geocenter relative to the SSB at the moment of Space Shuttle \textit{Discovery}'s launch with HST (April 24, 1990, 12:33:51 UTC). Compared to more up-to-date methods (described below), using \texttt{odelaytime}'s outdated \texttt{de200.fits} results in an offset in the predicted position of Earth of a few thousand kilometers with associated time biases of several milliseconds. Querying JPL Horizons, querying Astropy, or using a more up-to-date fits tabulation of Earth's position based on the JPL's ``DE440" ephemeris all agree to within a few kilometers. 

The JPL Horizons System is the standard tool for Solar System ephemeris calculations and is used by most modern interplanetary missions. The system can be accessed by a variety of means, including online and through the command-line. For our barycentric corrections here, we use \texttt{astroquery.jplhorizons}. Importantly, JPL Horizons element and vector queries from \texttt{astroquery} expect TDB timescales, while ephemeris queries expect UTC.\footnote{\url{https://stackoverflow.com/questions/67271676/astropy-and-jpls-horizons-query-have-different-output}} Querying with the wrong times can result in a time offset of over 69 seconds (the TDB-UTC difference), corresponding to offsets in the position of Earth of thousands of kilometers (as shown in Figure~\ref{fig:earth}), and inaccurate barycentric corrections at the millisecond level. Providing \texttt{astroquery} with an \texttt{astropy.time} object with the appropriate scale attribute defined does not mitigate this; one must provide \texttt{astroquery.jplhorizons} with the correct timescale values manually.

\texttt{Astropy} also has ephemeris services via \texttt{astropy.coordinates}, including a built-in Solar System ephemeris. Figure~\ref{fig:earth} shows implementations of \texttt{astropy.coordinates.get\_body\_barycentric} using two different ephemerides. We note that setting the ephemeris to ``jpl" (\texttt{solar\_system\_ephemeris.set('jpl')} chooses the DE430 ephemeris, created in 2013, while JPL Horizons will provide calculations based on DE440, created in 2020. Figure~\ref{fig:earth} shows a minor but non-zero difference of up to about a hundred kilometers due to this difference, which can result in offset barycentric corrections of a few tenths of a millisecond. \texttt{Astropy} users can select the ``DE440" ephemeris with \texttt{astropy.coordinates.solar\_system\_ephemeris.set('DE440')}.

\ssubsection{Calculating HST's Position}\label{sec:HSTpos}

\begin{figure}[h]
    \centering
    \includegraphics[width=0.9\linewidth]{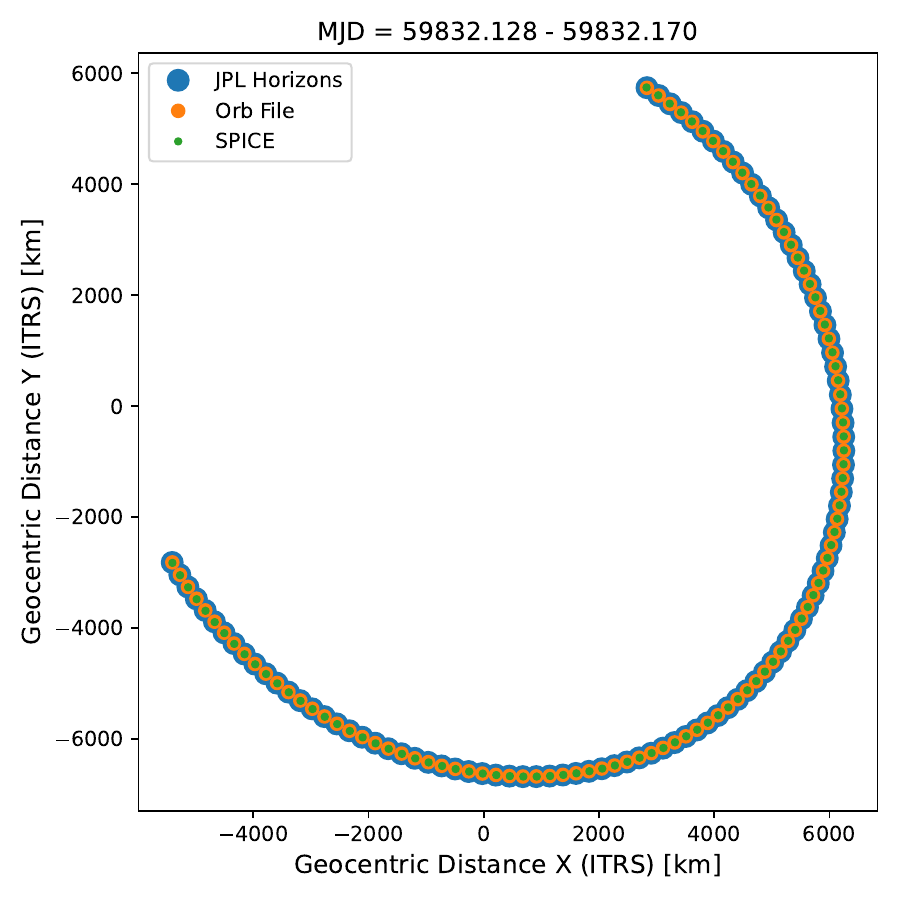}
    \vspace{-25pt}
    \caption{The position of HST over the course of 1 hour on MJD = 59832 in the IRTS reference system as queried from three difference sources. All sources, JPL Horizons, MAST HST orbital files, and the HST SPICE kernel agree to within a few km.}
    \label{fig:HST}
\end{figure}

Once the Earth's position relative to the SSB is known, the next step in calculating barycentric corrections is to determine the position of HST relative to the Earth. At an orbital altitude of about 475 km, the location of HST can produce an additional time delay of up to 46 milliseconds compared to if HST was on the other side of the Earth. This additional time delay will vary throughout HST orbits, such that the barycentric correction at the beginning of an HST orbit can be different than the correction at the end of an orbit. Therefore, our new tool must not only convert the exposure start and stop times in the file headers, but also every time recorded in \texttt{TIME-TAG} mode.

Since launch, HST's position has been recorded in orbital ephemeris files stored in the HST archive.\footnote{\url{https://www.stsci.edu/~STIS/monitors/ephemeris_files.html}} These files record HST's position and velocity relative to the geocenter in GCRS coordinates every five minutes over the course of 2-3 days. The IRAF \texttt{odelaytime} task used these files to interpolate HST's position at a given \texttt{TIME-TAG} timestamp. Retrieval of these files requires manual download via FTP, as these files are not available on the regular MAST portal. With FTP use deprecated in modern browsers, users can use the built-in Python \texttt{ftplib} or client apps like Cyberduck\footnote{\url{https://cyberduck.io/}} to access these files. Like \texttt{odelaytime}, \texttt{stistools.barycentric\_correction} is built to handle these HST ephemeris files.

In \texttt{stistools.barycentric\_correction} we also implement a more modern approach by querying JPL Horizons, similar to how we compute Earth's position. As with  the orbital ephemeris files, JPL Horizons uses GCRS to define the coordinates of HST's position. As an extra check, we directly loaded the SPICE (Spacecraft, Planet, Instrument, C-matrix, Events) kernel files with \texttt{SpiceyPy} and calculate HST's position directly. SPICE kernels are the underlying data sets of navigation and other information required for precision observation geometry\footnote{\url{https://naif.jpl.nasa.gov/naif/spiceconcept.html}}. Figure~\ref{fig:HST} shows the level of agreement between each method, which is on the order of a few kilometers.

We note that the HST position cannot be propagated into the future to high accuracy so observers should wait for the relevant trajectory data to become available before calculating barycentric corrections. It usually only takes a few days for new HST orbital files to become available. For JPL Horizons, it is less obvious whether up-to-date data is used because the position of the spacecraft is propagated forward in time using the latest trajectory data for planning purposes. Users can contact\footnote{\url{https://ssd.jpl.nasa.gov/horizons/manual.html\#ack}} JPL Solar System Dynamics to verify a spacecraft trajectory or request an update.

\begin{figure}[h]
    \centering
    \includegraphics[width=0.55\linewidth]{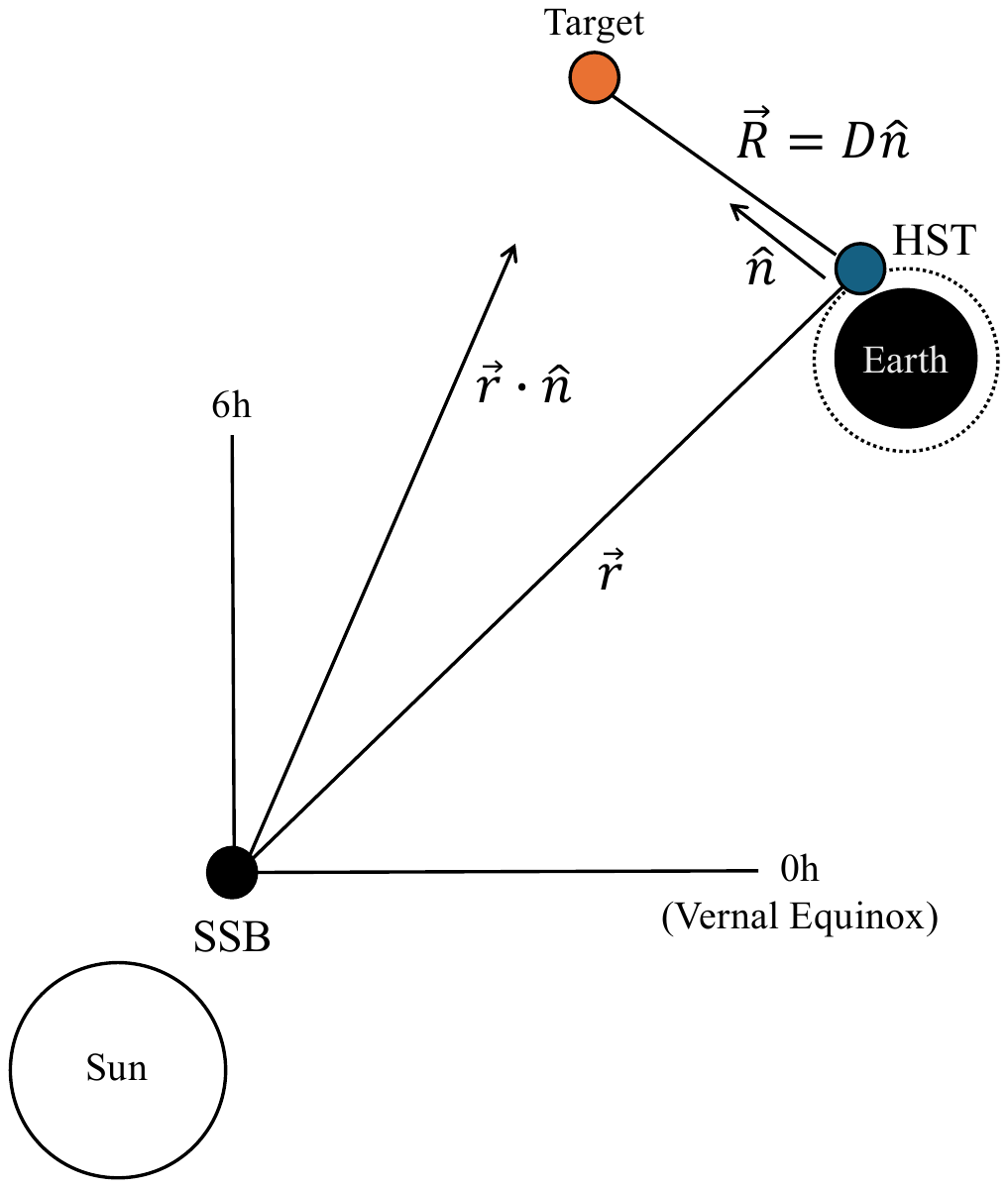}
    \caption{Geometry of the barycentric correction problem. The SSB's position relative to the Sun in 2026 is shown to scale. All other distances are not to scale.}
    \label{fig:geometry}
\end{figure}

We also note that Astropy builds URLs to query JPL Horizons, so queries need to be both be as small as possible and as infrequent as possible so as not to overload JPL servers. To that end, if multiple times are queried, \texttt{barycentric\_correction} requests times at a resolution of 1 minute, plus 5 minutes of padding at the beginning and end of observations to aid in interpolation. If a single time is queried, the exact time is used.

\ssection{Calculating Light Travel Times}\label{sec:LTT}

With known positions for both the Earth and HST, we can now calculate a barycentric correction for any part of the sky. As described in Eastman et al.~(2010), the conversion to our preferred BJD$_{\rm TDB}$ from JD$_{\rm UTC}$ is given by:

\begin{equation}
    {\rm BJD_{\rm TDB}} = {\rm JD_{\rm UTC}} + \Delta_{R\odot} + \Delta_C + \Delta_{S\odot} + \Delta_{E\odot}, \label{eq:tdb}
\end{equation}

\noindent where $\Delta_{R\odot}$ is the geometric R{\o}mer delay, $\Delta_C$ is the clock correction required to convert between the UTC and TDB time standard (e.g., leap seconds), $\Delta_{S\odot}$ is the Shapiro delay from light passing near a massive object, and $\Delta_{E\odot}$ is the general relativistic Einstein delay difference from our location relative to the geocenter.

The geometric R{\o}mer delay due to the finite speed of light is the largest term in the conversion and following Eastman et al.~(2010) can be represented to first order with the plane-wave approximation as 
\begin{equation}
    \Delta_{R\odot} = \frac{\overrightarrow{r} \cdot \hat{n}}{c},
\end{equation}

\noindent where $c$ is the speed of light, $\overrightarrow{r}$ is the vector from the origin (i.e., the SSB) to the observer (i.e., HST) and $\hat{n}$ is the unit vector from the observer to the object being observed. The geometry of these vectors is shown in Figure~\ref{fig:geometry}. For a given RA and DEC this can be expressed as

\begin{equation}
\left(
\begin{array}{c}
\cos(\mathrm{DEC})\cos(\mathrm{RA}) \\
\cos(\mathrm{DEC})\sin(\mathrm{RA}) \\
\sin(\mathrm{DEC})
\end{array}
\right).
\end{equation}

However, because this assumes the plane wave approximation, an additional second-order ``finite-distance" correction must be included if the object being observed is relatively nearby (i.e., inside the Solar System). This is fundamentally due to the curvature of the wavefronts, which can be assumed to be flat if the object is at infinity. This correction is \textit{not} included in the calculation of time delay in \texttt{astropy.Time.light\_travel\_time}, but can be relevant to HST users and was included in the original \texttt{odelaytime} task. We include it here in \texttt{barycentric\_correction} with the following equation:

\begin{equation}
    \frac{r^2-(\hat{n}\cdot\overrightarrow{r})^2}{2cD}\label{eq:FDC}
\end{equation}

\noindent where $D$ is the distance to the target, given by the user. With this correction factor, we find accuracies of about 5 ms inside the Solar System.

To perform the corrections in Equation~\ref{eq:tdb} within \texttt{barycentric\_correction}, we define a time object with the relevant format and scale (i.e., MJD$_{\rm UTC}$) along with a location relative to the SSB, found and verified as described in Sections~\ref{sec:Earthpos} and \ref{sec:HSTpos}. We then convert the time to the TDB time-standard when querying HST's position with JPL Horizons, adding the correct number of leap seconds and the 32.184 s clock offset. Next, we must define an \texttt{astropy.coordinates} object with our observed target. We do this based on the \texttt{RA\_TARG} and \texttt{DEC\_TARG} keywords from the fits file's primary header corresponding to the target position at the start of the observations. From this, we can calculate the light travel time to the barycenter with \texttt{astropy.Time.light\_travel\_time}, adding the finite distance correction in Equation~\ref{eq:FDC}. 

After adding the light travel time calculated above to the times in the fits data files, we now have a time in BJD$_{\rm UTC}$. At this point, \texttt{odelaytime} would add a $\approx1.34$ms gravitational time delay term (using a set of orbital elements for Earth calculated in 1981), making a time with no common system definition -- the accounting for both R{\o}mer and Einstein delay would take it two steps towards BJD$_{\rm TDB}$ in Equation~\ref{eq:tdb}, but without the clock corrections, $\Delta_C$, it would still be off from BJD$_{\rm TDB}$ by the number of leap seconds in UTC plus the 32.184 second TAI-TT offset. 

Our final step is to apply the clock corrections ($\Delta_C$ in Eq.~\ref{eq:tdb}) to convert the times from BJD$_{\rm UTC}$ to BJD$_{\rm TDB}$ with astropy. Users can choose to keep their times in BJD$_{\rm UTC}$, however. Afterwards, the corrected times are changed in the fits files, the selected time system is written in the fits header history, and \texttt{DLAYCOR = COMPLETE} is added as a header keyword to prevent double-correcting the times.

\section{Verification of \texttt{barycentric\_correction}}

We used a variety of tools to verify the accuracy of \texttt{barycentric\_correction}. Table~\ref{tab:example} shows an example of the clock corrections for the exposure start time of observations of exoplanet WASP-121b from dataset OD9M97020 (PI: Sing). Given an original time from the pipeline (MJD$_{\rm UTC}$ = 57808.01067964), we show the resulting barycentric corrections. Initially, we created a direct Python port of the IRAF \texttt{odelaytime} routine, but subsequent testing showed additional improvements were necessary, like the updating of the Earth ephemeris file and self-contained HST orbital position querying. For example, Table~\ref{tab:example} shows a $>$3~ms disagreement from the \texttt{odelaytime} port when using the original \texttt{odelaytime} Earth ephemeris file based on DE200 compared to the new ephemeris based on DE440.

The times calculated from the \texttt{odelaytime} port with the updated Earth ephemeris file and \texttt{barycentric.correction} in BJD$_{\rm UTC}$ show agreement at the 1 ms level. When accounting for the difference in treatment of the Einstein delay term as noted in Section~\ref{sec:LTT}, both agree to better than 1 ms. Notably, both tools are using different sources for Earth's position relative to the SSB, different sources for HST's position, and different routines to calculate the corrections themselves.

We also compared corrections using \texttt{barycentric.correction} in BJD$_{\rm TDB}$ with JPL Horizons versus the HST orbital files to calculate HST's position relative to the geocenter. Again, both routines agree within 1 ms. We note that the difference between the corrections in BJD$_{\rm UTC}$ and BJD$_{\rm TDB}$ is the expected 69.184~s from the clock corrections (Eastman et al.~2010), with 32.184 s being the offset between International Atomic Time and Terrestrial Time (the latter from which TDB is derived) and 37 seconds from leap seconds between International Atomic Time and UTC.

As a third-party check, we also compare to \texttt{barycorrpy.utc\_tdb} from Kanodia et al.~(2018), again finding agreement better than 1 ms with \texttt{barycentric.correction}. Throughout testing, we also verified geocentric barycentric corrections with \texttt{pintbary}  (Luo et al.~2019, Susobhanan et al.~2024) and \texttt{barycorr}.\footnote{\url{astroutils.astronomy.osu.edu} (now defunct).}

\begin{table}[h!]
    \centering
    \begin{threeparttable}
\caption{Time Correction Comparison (Dataset OD9M97020) \label{tab:example}}
\begin{tabular}{lcccc}
\hline
\textbf{Tool} & \textbf{HST Position} & \textbf{Time Scale} & \textbf{Time} (MJD-57808) & \textbf{$\Delta$ Time} (s) \\
\hline
Pipeline & --  & MJD$_{\rm UTC}$     & 0.01067964 & -- \\
\texttt{odelaytime} port\tnote{a}  & orbfile    &  BJD$_{\rm UTC^*}$\tnote{c} & 0.012926078 & 194.0922 \\
\texttt{odelaytime} port\tnote{b}  & orbfile    &  BJD$_{\rm UTC^*}$\tnote{c} & 0.012926040 & 194.0889 \\
\texttt{barycentric\_correction} &JPL Horizons &  BJD$_{\rm UTC}$ & 0.013726789 & 194.0876 \\
\texttt{barycentric\_correction} &  JPL Horizons &    BJD$_{\rm TDB}$   & 0.013726780 & 263.2729 \\
\texttt{barycentric\_correction} & orbfile  &    BJD$_{\rm TDB}$   & 0.012926024 & 263.2736 \\

\texttt{barycorrpy.utc\_tdb} & orbfile  &    BJD$_{\rm TDB}$   & 0.013726780 & 263.2729\\

\hline
    \end{tabular}
    \begin{tablenotes}
        \item[a] Using the old Earth ephemeris file based on DE200.
        \item[b] Using the new Earth ephemeris file based on DE440.
        \item[c] BJD$_{\rm UTC^*}$ refers to BJD$_{\rm UTC}$ plus the $\approx1.34$ms Einstein delay term added by \texttt{odelaytime}.
    \end{tablenotes}
    \end{threeparttable}
\end{table}

\ssection{Absolute Time Accuracy}

So far, we have focused solely on the issues related to transforming HST observation times to those described by a standardized observing position in a specific time system, accounting for the light travel time and general relativistic effects to arrive at highly precise times. However, the accuracy of these precise corrections also depend on the absolute accuracy of the time stamps HST places on the observed events.

The \textit{Hubble Space} Telescope and its instruments use several onboard clocks to mark the passage of time. In general, these consist of elapsed time counters that mark the number of ticks from some zero point. HST's main computer counts every 125 ms in a 32-bit counter, rolling over every 17 years. All clocks will drift with respect to their initial zero point, so times aboard HST are now correlated with UTC within a mission-required 10 ms. 

However, each of the instruments have their own clock for use with their internal flight software. For STIS, the timestamps for the data headers and \texttt{TIME-TAG} events are built from two times. First, a 16-bit 82C54 hardware timer decrements starting from 255 every 125 microseconds, setting the time-resolution of the timestamps. When this ``fine" timer reaches zero from 255 after 32 ms, it sends an interrupt signal to the MAMA Interface Electronics (MIE) that increments a 31-bit coarse timer.

When a photon is detected by the MAMA detectors, the X,Y centroid is inserted in a first-in first-out (FIFO) queue. A process on the MIE computer extracts events from the queue and writes a  32-bit time-tag event word (which contains the current fine time and the extracted X, Y centroid) into the STIS buffer. At the start of the process, the MIE first checks whether the STIS coarse time has changed since the last time the coarse time was written to the STIS memory buffer. If so, a new coarse time word is written to the STIS memory buffer before the \texttt{TIME-TAG} event word is written.  The entire event processing time is short enough that 4 events can be processed during each 125 microsecond tick of the fine timer. Therefore, at global detector count rates of 32,000 ct/s or less, the times that events enter and leave the FIFO queue are identical within the precision of fine timer. Above 32,000 ct/s, events must wait in the queue to be processed. Since the time written in the \texttt{TIME-TAG} event word is the time the event is \textit{processed}, events can accumulate delays of up to 82 milliseconds depending on how many events were waiting in the queue when it was added. See STIS IHB (\S 11.12.3) for additional information on \texttt{TIME-TAG} event processing at high count rates.

The coarse time is reset and synced with the \textit{Hubble Space Telescope} clock during the monthly anneal of the CCD detector. The coarse time is otherwise not expected to drift much over the course of this month. Thus, the expected accuracy of the STIS timestamps (below count rates of 32,000 ct/s) is the combined accuracy of the HST clock ($\leq$10 ms) and the resolution of the fine time ($\leq$125 microseconds). We therefore expect all STIS timestamps to be accurate to 10.125 ms with a precision of 125 microseconds.

\ssection{Summary}\label{sec:conclusion}

We have presented a new \texttt{stistools} utility, \texttt{barycentric\_correction}, to calculate barycentric corrections for HST observations. The script is designed to work with STIS and COS files of all kinds (e.g., tag, raw, flt, x1d, and x2d files). \texttt{barycentric\_correction} uses JPL Horizons and \texttt{Astropy} to implement modern position determination and time-delay calculations. The script has been tested to ms accuracy for objects outside the Solar System and to about 5 ms for objects inside the Solar System (where the plane-wave approximation breaks down).




\vspace{-0.3cm}
\ssectionstar{Acknowledgements}
\vspace{-0.3cm}

We thank John Debes for a constructive review of the initial version of this ISR. We thank Steve Arslanian at Goddard Space Flight Center for helpful discussion of the onboard clock hardware and software.

\vspace{-0.3cm}
\ssectionstar{Change History for STIS ISR 2026-02}\label{sec:History}
\vspace{-0.3cm}
Version 1: \ddmonthyyyy{28 May 2026} - Original Document 

\vspace{-0.3cm}
\ssectionstar{References}\label{sec:References}
\vspace{-0.3cm}

\noindent
Carlot, P. et al., 2020, A\&A, 644, A159\\
Eastman, J. et al., 2010, PASP, 122, 894, 935\\
International Astronomical Union, Resolution A4 (1991), \url{www.iau.org}\\
Kanodia, S. \& Wright, J.~T., 2018, ASCL, record ascl:1808.001, \url{https://github.com/shbhuk/barycorrpy}\\
Lorimer, D.~R., 2008, Living Reviews in Relativity, 11, 1, 8\\
Luo, J. et al. 2019, ASCL, ascl:1902.007 \\
Susobhanan, A. et al. 2024, ApJ, 971, 2 150.\\
Valenti, J.~A. et al., 2008, STIS Technical Instrument Report, 2008-01\\
Wilkins, G. A., 1990, IAU Symposium, 141, 39-46\\

\end{document}